\documentclass{article}

\usepackage{arxiv}

\usepackage[utf8]{inputenc} % allow utf-8 input
\usepackage[T1]{fontenc}    % use 8-bit T1 fonts
\usepackage[colorlinks,bookmarksopen,bookmarksnumbered,citecolor=red,urlcolor=black]{hyperref}
\usepackage{url}            % simple URL typesetting
\usepackage{booktabs}       % professional-quality tables
\usepackage{amsfonts}       % blackboard math symbols
\usepackage{nicefrac}       % compact symbols for 1/2, etc.
\usepackage{microtype}      % microtypography
\usepackage{graphicx}
\usepackage[super]{natbib}
\usepackage{doi}
\usepackage{moreverb,url}
\usepackage{scalefnt}
\usepackage{caption}
\usepackage{mathrsfs}
\usepackage{amsmath}
\usepackage[mathscr]{euscript}
\usepackage{rotating}
\usepackage{float}
\usepackage{bm}
\usepackage{multirow}
\usepackage{multicol}
\usepackage[all]{xy}
\usepackage{tikz}
\usetikzlibrary{shapes.geometric, arrows}

\title{A Bayesian survival model induced by hurdle zero-modified power series discrete frailty with dispersion: an application in lung cancer}

\date{} 					% Or removing it

\author{ \href{https://orcid.org/0000-0002-2614-2336}{\includegraphics[scale=0.06]{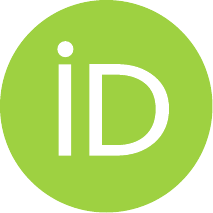}\hspace{1mm}Katy C.~Molina} \\
	\small Institute of Mathematics and Computer Sciences\\
	\small University of S{\~a}o Paulo\\
	\small S{\~a}o Carlos, Brazil \\
	\small and \\
	\small Dept. of Statistics\\
	\small Federal University of S{\~a}o Carlos\\
	\small S{\~a}o Carlos, Brazil \\
	\small \texttt{rocio.cm@usp.br} \\
	\And
	\href{https://orcid.org/0000-0002-1016-8734}{\includegraphics[scale=0.06]{orcid.pdf}\hspace{1mm}Joaqu{\'i}n~Mart{\'i}nez-Minaya} \\
	\small Dept. of Applied Statistics and Operational Research and Quality\\
	\small Universitat Polit{\`e}cnica de Val{\`e}ncia\\
	\small Valencia, Spain \\
	\small \texttt{jmarmin@eio.upv.es} \\
	\And
	\href{https://orcid.org/0000-0003-3764-0397}{\includegraphics[scale=0.06]{orcid.pdf}\hspace{1mm}Danilo~Alvares} \\
	\small MRC Biostatistics Unit\\
	\small University of Cambridge\\
	\small Cambridge, UK \\
	\small \texttt{danilo.alvares@mrc-bsu.cam.ac.uk} \\
        \phantom{\small Institute of Mathematics and Computer Sciences}
	\And
	\href{https://orcid.org/0000-0002-6780-2089}{\includegraphics[scale=0.06]{orcid.pdf}\hspace{1mm}Vera D.~Tomazella} \\
	\small Dept. of Statistics\\
	\small Federal University of S{\~a}o Carlos\\
	\small S{\~a}o Carlos, Brazil \\
	\small \texttt{vera@ufscar.br} \\
	\phantom{\small Dept. of Applied Statistics and Operational Research and Quality}
}

\date{}

\renewcommand{\shorttitle}{A Bayesian survival model induced by HZMPS discrete frailty with dispersion}

\hypersetup{
pdftitle={A Bayesian survival model induced by hurdle zero-modified power series discrete frailty with dispersion: an application in lung cancer},
pdfauthor={Katy C.~Molina, Joaqu{\'i}n~Mart{\'i}nez-Minaya, Danilo~Alvares, Vera D.~Tomazella},
pdfkeywords={Bayesian inference, Frailty survival models, Hurdle reparameterization, Zero-modification specification},
}

\begin{document}
\maketitle

\begin{abstract}
Frailty survival models are widely used to capture unobserved heterogeneity among individuals in clinical and epidemiological research. This paper introduces a Bayesian survival model that features discrete frailty induced by the hurdle zero-modified power series (HZMPS) distribution. A key characteristic of HZMPS is the inclusion of a dispersion parameter, enhancing flexibility in capturing diverse heterogeneity patterns. Furthermore, this frailty specification allows the model to distinguish individuals with higher susceptibility to the event of interest from those potentially cured or no longer at risk. We employ a Bayesian framework for parameter estimation, enabling the incorporation of prior information and robust inference, even with limited data. A simulation study is performed to explore the limits of the model. Our proposal is also applied to a lung cancer study, in which patient variability plays a crucial role in disease progression and treatment response. The findings of this study highlight the importance of more flexible frailty models in survival data analysis and emphasize the potential of the Bayesian approach to modeling heterogeneity in biomedical studies.
\end{abstract}


%%%%%%%%%%%%%%%%%%%%%%%%%%%%%%%%%%%%%%%
\section{Introduction} \label{Section:introduction}

Survival analysis provides the fundamental statistical framework for modeling data where the time until an event of interest occurs is the key variable. Its importance transcends a single discipline with crucial applications in diverse fields. In medicine and epidemiology, it is indispensable for studying disease progression and evaluating treatment effectiveness.\cite{Collett2015} In reliability engineering, it models the operational lifetime or failure time of systems and components.\cite{Meeker2022} In actuarial science and economics, it facilitates the prediction of longevity\cite{Dickson2020} and the modeling of financial and credit risks. More recently, these techniques have also been integrated into machine learning to effectively handle censored data within predictive models.\cite{Wang2019}

In cancer studies, survival analysis allows for a better understanding of factors that influence patient survival, evaluation of treatment effectiveness, and identification of potential prognostic variables. Survival models are widely used to analyze the impact of different clinical and genetic characteristics on disease progression, assisting in medical decision-making and the development of more effective therapeutic strategies.

According to the World Health Organization, \cite{WHO2024Cancer} cancer remains one of the leading causes of death worldwide, responsible for approximately 19.3 million new cancer cases and nearly 10 million cancer deaths in 2020. Despite its global impact, advances in cancer research, early detection and the development of new therapies have significantly improved survival rates in many types of cancer. For instance, cancers such as those originating in the breast, prostate, and thyroid, along with melanoma, often exhibit high cure rates, particularly when detected at early stages due to effective treatment modalities. Conversely, other malignancies present formidable therapeutic challenges. Pancreatic, lung, and liver cancers are frequently associated with poorer prognoses, often attributed to aggressive biological behavior, late-stage diagnosis, or limited treatment efficacy.\cite{Siegel2024} 

Among the malignancies presenting significant therapeutic challenges, lung cancer is particularly prominent due to its high mortality burden. Indeed, it stands as the leading cause of cancer-related death worldwide, a status underscored by data from GLOBOCAN 2020. These estimates, compiled by the International Agency for Research on Cancer and reported in Ferlay et al.\cite{ferlay2024global}, indicate that lung cancer accounted for approximately 1.8 million deaths in that year. 

The effectiveness of treatment varies according to several factors, such as the type of cancer, the stage of diagnosis, and the individual characteristics of each patient. In this context, the study patients are different from each other; e.g., they have different medical outcomes, lifestyles, different reactions to medications, etc. This diversity, also referred to as heterogeneity, is generally not considered as it is difficult to quantify, yet its inclusion is crucial for accurately modeling patient outcomes. Vaupel et al.\cite{vaupel1979impact} were the pioneers in the development of a model that included the presence of a random effect, also called frailty term, representing unobserved factors containing information that could not be considered during the study. Frequently, this frailty term is included multiplicatively in the hazard function and consequently becomes a generalization of the risk model provided by Sir David Cox\cite{cox_oakes}. The natural candidates for the distribution of frailty are nonnegative and continuous. However, the use of a continuous distribution for frailty does not allow the existence of individuals with zero frailty. In this way, the use of a discrete distribution for frailty allows one to obtain a proportion of patients with zero frailty, who consequently are long-term survivors. 

Frailty modeling using discrete distributions has been explored by several authors. Wienke\cite{wienke2010frailty} used a discrete Poisson process for this purpose. Caroni et al.\cite{caroni2010} investigated Poisson, negative binomial, and geometric distributions as candidates in this context. Ata and {\"O}zel\cite{ata2013survival} developed a frailty model also based on a discrete Poisson process. Additionally, Sousa et al.\cite{de2017bayesian} proposed the hyper-Poisson distribution to model survival associated with such frailty. Cancho et al.\cite{cancho2019new} investigated zero-inflated power series (ZIPS). More recently, Molina et al.\cite{molina2021survival} incorporated the zero-modified power series (ZMPS) family distribution as a flexible alternative to model unobserved heterogeneity. Cancho et al.\cite{cancho2021bayesian} used the mixed Poisson distribution and Espirito Santo et al.\cite{do2022survival} worked with the Kats distribution. These approaches naturally align with the class of cure rate models introduced by Tsodikov et al.\cite{tsodikov2003}, wherein a fraction of the population is considered to be non-susceptible to the event of interest. Consequently, these frailty models can be regarded as particular formulations within the general framework of cure rate modeling, offering an effective mechanism to capture the presence of long-term survivors through a discrete frailty structure.

In this paper, we propose the use of the hurdle zero-modified power series (HZMPS) family of discrete distributions for modeling the frailty term. Incorporating a dispersion parameter, these distributions offer greater flexibility than ZMPS distributions, previously studied by Molina et al.\cite{molina2021survival}. The main advantage of using this family of distributions in the frailty specification is that it does not require prior knowledge about the specific behavior of the survival data. The model can identify the presence or absence of long-term survivors, adjusting to particular cases according to the frequency of zero in the data (zero-inflated, zero-deflated, zero-truncated, or traditional power series distribution).

This paper is organized as follows. Section \ref{Section:zmps_hurdle} introduces and discusses the HZMPS distribution. Section \ref{Section:frailtymodel_zmgp} presents the new proposed frailty model. Section \ref{Section:bayesian_inference} details the new Bayesian survival model fitted using the hurdle zero-modified generalized Poisson (HZMGP) distribution. Simulation studies are presented in Section \ref{Section:simulation}. In Section \ref{Section:application}, we demonstrate the applicability of the proposed model with a real-world lung cancer study. Finally, Section \ref{Section:finalremarks} offers concluding remarks. The HZMGP-based model was implemented using R-Stan, \cite{carpenter2017stan} and the code is available at \url{www.github.com/Katy-RCM/HZMGP}.
%%%%%%%%%%%%%%%%%%%%%%%%%%%%%%%%%%%%%%%

%%%%%%%%%%%%%%%%%%%%%%%%%%%%%%%%%%%%%%%
\section{HZMPS distribution} \label{Section:zmps_hurdle}

In many real statistical problems, the observed datasets are natural values, commonly called counts. In this sense, due to the characteristics of the data, the Poisson distribution has been widely employed in different studies.\cite{frome1983analysis,brown2002test,hayat2014understanding,inouye2017review,kurnia2021analysis} However, the use of this distribution in count data analysis does not guarantee a good fit, as its applicability is limited by the assumption that the mean and variance are equal. When the mean value is less than the variance, we call this overdispersion and the opposite case is called underdispersion. More flexible alternatives for modeling count data include the negative binomial and the generalized Poisson (GP) distributions. Despite this flexibility, both may still fall short in scenarios with an excess of zeros.\cite{kamalja2018estimation,kurnia2021analysis}

To address this issue, Concei{\c{c}}{\~a}o et al.\cite{conceicao2017} proposed the family of zero-modified power series (ZMPS) distributions which introduce a flexible framework for modeling count data with varying degrees of zero inflation or deflation. Specifically, let $Y$ be a random variable following a ZMPS distribution, $Y \sim \mbox{ZMPS}(\rho,\mu,\phi)$, where $\rho$, $\mu$ and $\phi$ represent zero-modification probability, mean and dispersion, respectively, and its probability mass function (p.m.f.) is defined as follows
\begin{equation}
\pi_{_\text{ZMPS}}(Y=y; \, \rho,\mu,\phi) = (1-\rho) \mathbb{I}(y) + \rho \, \pi_{_\text{PS}}(Y= y;\, \mu,\phi), \quad y \in \mathcal{A}_{0},
\label{distZMPS}
\end{equation}  

\noindent being ${\mathcal{A}}_{0} = \left\{0,1,2,\ldots\right\}$ the support formed by the subset of non-negative integers, $\pi_{_\text{PS}}(Y=y; \, \mu,\phi) $ is the power series (PS) distribution and $\mathbb{I}(y)$ an indicator function such that
\begin{equation*}
\mathbb{I}(y)=
\begin{cases}
1, \text{ se $y=0$},\\
0, \text{ se $y>0$}.
\end{cases}
\end{equation*} 

The restriction of the parameter $\rho$ carry out the following condition
\begin{equation}
0 \le \rho \le \frac{1}{1-\pi_{_\text{PS}}(Y=0; \, \mu,\phi)},
\label{rest_p}
\end{equation}

\noindent where $\pi_{_\text{PS}}(Y=0; \, \mu,\phi)$ is the PS distribution evaluated at zero. 

Concei{\c{c}}{\~a}o et al.\cite{conceicao2017} also proposed a reparameterization for $\rho$, which it is useful in different situations. Specifically, Equation \eqref{distZMPS} can be rewritten as
\begin{align}
\pi_{_\text{ZMPS}}(Y =y; \, \rho,\mu,\phi) &= (1-\rho) \mathbb{I}(y) +\rho \, \pi_{_\text{PS}}(Y=y; \, \mu,\phi), \quad y \in \mathcal{A}_0 \nonumber \\
&= \underbrace{1 - \rho + \rho \, \pi_{_\text{PS}}(Y=0; \, \mu,\phi)}_{\substack{y = 0}} + 
\underbrace{\rho \, \pi_{_\text{PS}}(Y=y; \, \mu,\phi)  }_{\substack{y > 0}} \nonumber \\
&= \pi_{_\text{ZMPS}}(Y=0; \, \rho,\mu,\phi) \mathbb{I}(y) + \pi_{_\text{ZMPS}}(Y=y; \, \rho,\mu,\phi)(1 - \mathbb{I}(y)) \nonumber \\
&= \big[ 1- \rho + \rho \, \pi_{_\text{PS}}(Y=0; \, \mu,\phi) \big] \mathbb{I}(y) + \rho \, \pi_{_\text{PS}}(Y=y; \, \mu,\phi)(1 - \mathbb{I}(y)). 
\label{zmps_repa}
\end{align}

The restriction of $\rho$ in \eqref{rest_p} can be rewritten as
\begin{align}
0 &\le \rho \, [ 1-{\pi}_{_\text{PS}}(Y=0; \, \mu,\phi)] \le 1  \nonumber \\
0 & \le \omega \le 1.
\label{restri}
\end{align}

Now, by considering \eqref{zmps_repa} subject to the restriction \eqref{restri}, $Y \sim \mbox{HZMPS}(\omega,\mu,\phi)$ and its p.m.f. is given by
\begin{equation}
\pi_{_\text{HZMPS}}(Y=y; \, \omega,\mu,\phi) = (1-\omega) \mathbb{I}(y) + \omega \, \pi_{_\text{ZTPS}}(Y=y; \, \mu,\phi), \quad y \in \mathcal{A}_0,
\label{distZMPS_repa}
\end{equation} 

\noindent where $\pi_{_\text{ZTPS}}(Y=y; \, \mu,\phi) = \left\{\displaystyle\frac{\pi_{_\text{PS}}(Y=y; \, \mu,\phi)}{1-\pi_{_\text{PS}}(Y=0; \, \mu,\phi)}\right\} \mathbb{I}_{\mathcal{A}_1}(y)$ is the zero-truncated power series (ZTPS) distribution, and its support is $\mathcal{A}_1 = \{1,2,3,\ldots\} \subset \mathcal{A}_0$. Also, 
$\mathbb{I}_{\mathcal{A}_1}(y) = 1 - \mathbb{I}(y)$ is an indicator function such that $\mathbb{I}_{\mathcal{A}_1}(y)=1$ if $y \in \mathcal{A}_1$ and $0$ if $y \notin \mathcal{A}_1$.

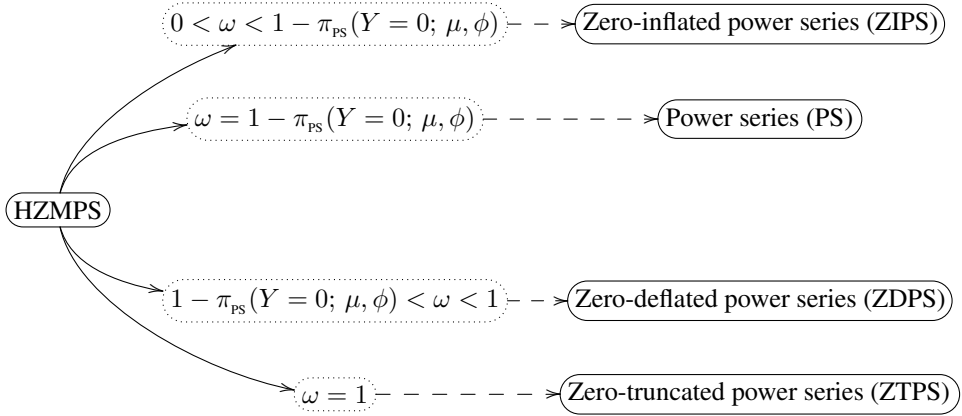
\begin{figure}[ht!]
\centering
\scalefont{1}
\begin{center}
\xymatrix @=.7cm{& *+[F.:<3.5mm>] \txt{$0<\omega<1 - \pi_{_\text{PS}}(Y=0; \, \mu,\phi)$} \ar@{-->}[r] & *+[F:<3.5mm>]\txt{\parbox{4.7cm}{Zero-inflated power series (ZIPS)}}
\\
& *+[F.:<3.5mm>]\txt{$\omega = 1 - \pi_{_\text{PS}}(Y=0; \, \mu,\phi)$} \ar@{-->}[r] & *+[F:<3.5mm>]\txt{\parbox{2.5cm}{Power series (PS)}} \\
*+[F:<3.5mm>]\txt{HZMPS} \ar@(u,l)[uur] \ar@(u,l)[ur] \ar@(d,l)[dr] \ar@(d,l)[ddr]\\
& *+[F.:<3.5mm>]\txt{$1 - \pi_{_\text{PS}}(Y=0; \, \mu,\phi) < \omega < 1$} \ar@{-->}[r] & *+[F:<3.5mm>]\txt{\parbox{4.9cm}{Zero-deflated power series (ZDPS)}} \\
& *+[F.:<4.5mm>]\txt{$\omega = 1$} \ar@{-->}[r] & *+[F:<4.5mm>]\txt{\parbox{5.1cm}{Zero-truncated power series (ZTPS)}} }
\end{center}
\caption{Particular cases for the hurdle zero-modified power series (HZMPS) distributions.}
\label{Figure:HZMPS}
\end{figure}

According to Concei{\c{c}}{\~a}o et al.\cite{conceicao2017}, the HZMPS distribution contains the ZTPS distribution as one of its components, which differs from the traditional mixture representation of zero-inflated distributions. Moreover, the HZMPS distribution can be interpreted as a superposition of two processes, i.e., one that produces only zero observations and another that produces only positive observations from a ZTPS distribution. One of the advantages of this reparametrization is that $\omega$ and $\mu$ are orthogonal, allowing us to estimate $\omega$ independently of $\mu$. However, the parametrization provided in \eqref{distZMPS} enables inference on $\rho$, which facilitates the direct identification of the type of zero-modification (zero-inflated, zero-deflated, zero-truncated, or traditional PS distribution).\cite{conceiccao2017biparametric} This interpretation holds with $\omega$, in reparametrization \eqref{distZMPS_repa}. Furthermore, particular cases are obtained according to the value of $\omega$ (see Figure \ref{Figure:HZMPS}).

Some of the distributions that belong to the HZMPS family are the hurdle zero-modified Poisson (HZMP), hurdle zero-modified geometric (HZMG), hurdle zero-modified binomial (HZMB), hurdle zero-modified generalized Poisson (HZMGP), hurdle zero-modified negative binomial (HZMNB), among others.  

The HZMPS family offers notable flexibility due to the inclusion of a dispersion parameter, which is also useful in situations presenting diverse behaviors regarding zero frequencies. Among the distributions available within this family, this paper focuses on the HZMGP distribution. Table~\ref{tabela:algumasdistri} details the main characteristics of the HZMGP distribution, including its p.m.f., probability generating function (p.g.f.), mean, and variance.

\begin{table}[ht!]
\centering
\caption{Description of some characteristics of HZMGP distribution.}
\label{tabela:algumasdistri}
\begin{tabular}{c|c}
\hline
Characterization & Formula \\
\hline
p.m.f. & $\left( 1-\omega  \right) \mathbb{I} \left( y \right) + \omega
\displaystyle\left[\frac{ \displaystyle\frac{(1+\phi y)^{y-1}}{y!} \left[\displaystyle\frac{\mu e^{-\frac{\mu\phi}{1+\mu\phi}}}{1+\mu\phi}\right]^{y} }{ e^{\frac{\mu}{1+\mu\phi}} \left( 1-e^{-\frac{\mu}{1+\mu\phi}} \right)  }\right], \quad y \in \mathcal{A}_0 $ \\
\hline
p.g.f. & $ 1-\displaystyle\frac { \omega}{ 1-{ e }^{ -\frac { \mu  }{ 1+\mu \phi  }  } }\left[1- { e }^{ -\frac { 1 }{ \phi  } \left[ \mathbb{W}\left( -\frac { \mu \phi  }{ 1+\mu \phi  } r { e }^{ -\frac { \mu \phi  }{ 1+\mu \phi  }  } \right) +\frac { \mu \phi  }{ 1+\mu \phi  }  \right]  } \right],\quad 0 \leq r \leq 1$ \\
\hline
Mean & $\displaystyle\frac { \omega \mu}{ 1-{ e }^{ -\frac { \mu  }{ 1+\mu \phi  }  }}$ \\
\hline
Variance & $\displaystyle\frac { \omega \mu}{ 1-{ e }^{ -\frac { \mu  }{ 1+\mu \phi  }  }} { \left( 1+\mu \phi  \right)  }^{ 2 }+ \displaystyle\frac { \omega { \mu  }^{ 2 }}{ 1-{ e }^{ -\frac { \mu  }{ 1+\mu \phi  }  }} { \left( 1-\displaystyle\frac { \omega  }{ 1-{ e }^{ -\frac { \mu  }{ 1+\mu \phi  }  }} \right)  } $ \\
\hline
\end{tabular}
\end{table}

The expression for p.g.f. is based on the formulation proposed by Ambagaspitiya and Balakrishnan\cite{ambagaspitiya1994compound}, where $\mathbb{W}(\cdot)$ is the Lambert W function defined as $\mathbb{W}(x)e^{\mathbb{W}(x)}=x$. For more details on Lambert's W function, see Corless et al.\cite{corless1996lambertw}. The HZMGP distribution reduces to HZMP when $\phi = 0$.

\section{Frailty model induced by HZMGP distribution} \label{Section:frailtymodel_zmgp}

Frailty models provide a useful approach to incorporate a random effect, also called frailty or unobserved heterogeneity in survival analysis. This latent variable, denoted by $V$, represents the impact of relevant risk factors that are often not fully measured or included in the study. The incorporation of frailty into the Cox's models is typically done multiplicatively, often within the framework of proportional hazards. This approach confers considerable flexibility to the standard Cox model. 
Essentially, the introduction of frailty allows each individual to manifest a distinct and inherent risk profile, often extending beyond the risk predicted by the observed covariates alone.\cite{wienke2010frailty}

According to the proportional hazard model described by Cox and Oakes\cite{cox_oakes}, the conditional hazard is expressed by
\begin{equation}
h(t \mid V)=Vh_0(t),
\label{fragilidade}
\end{equation}

\noindent where $h_0(\cdot)$ is the baseline hazard function. The conditional survival function is defined as
\begin{equation}
S\left(t \mid V\right)={ e }^{ -V \int _{ 0 }^{ t }{ { h }_{ 0 }\left( s \right) ds }  }={ \left[ { S }_{ 0 }\left( t \right)  \right]  }^{V},
\label{suv}
\end{equation}

\noindent where $S_{0}(\cdot)$  denotes the baseline survival function. 

The frailty term is typically assumed to be a non-negative, continuous, and time-invariant random variable. Among the most widely adopted distributions for modeling frailty are the gamma distribution,\cite{vaupel1979impact,hougaard1984life} the inverse Gaussian distribution,\cite{hougaard1984life} the log-normal distribution,\cite{dossantos1995} and the power variance function family,\cite{calsavara_pvf} among others. However, the expression in Equation \eqref{fragilidade} is an unrealistic situation in the context of long-term survivors, and in certain situations it is convenient to conceive of the idea that frailty is discretely distributed. According to Caroni et al.\cite{caroni2010}, this approach is particularly appropriate when lifetime heterogeneity arises from exposure to a random number of harmful events or from the occurrence of a random number of failures within a unit. So, we assume that $V$ has a discrete distribution over the non-negative integers.

Let $\mathbb{P}(V = v) = q_{v}$ be the probability distribution of $V$, for $v=0,1,\ldots$ and taking proportional hazards, the marginal survival function of $T$ is given by
\begin{equation}
S(t)=\displaystyle \sum_{v=0}^{\infty}\mathbb{P}(V=v) \, S_{0}(t)^{v} = \displaystyle \sum_{v=0}^{\infty}q_{v} \, S_{0}(t)^{v} = G_{V} \,{S}_{0}(t),
\label{sobdiscreta}
\end{equation}

\noindent where ${G}_{V}(\cdot)$ is the p.g.f. of the random variable $V$. The survival function in \eqref{sobdiscreta} can be rewritten as
\begin{equation*}
S(t) = q_{0}+\sum_{v=1}^{\infty} q_{v} \, S_{0}(t)^{v},
\end{equation*}

\noindent and zero frailty or also called long-term survivors is obtained when $q_0 = \displaystyle \lim_{t \rightarrow \infty} S(t)$.

We employ the HZMGP distribution (see Table \ref{tabela:algumasdistri}) as a candidate for the frailty term. Its key advantage lies in its flexibility, derived from the dispersion parameter, which is beneficial for capturing patterns of unobserved heterogeneity. The dispersion parameter might also offer interpretable information about frailty variability. The core functions defining this model are presented in Table \ref{tab:models}. We used the Weibull distribution to specify the baseline hazard function, $h_{0}(t) = \gamma \, \lambda^{\gamma} \, t^{\gamma-1}$, and baseline survival function, $S_{0}(t) = e^{-(\lambda t)^{\gamma}}$, for $t>0$, $\lambda>0$, and $\gamma>0$.

\begin{table}[h!]
\centering
\caption{Survival function $S(t)$, hazard function $h(t)$, density function $f(t)$, and long-term survivors $q_0$ for the HZMPG model.}
\label{tab:models}
\begin{tabular}{c|c}
\hline 
Function & HZMGP model \\
\hline 
$S(t)$ & $1-\displaystyle\frac{\omega}{1-e^{-\frac { \mu  }{ 1+\mu \phi  } }}+\displaystyle\frac{\omega}{1-e^{-\frac { \mu  }{ 1+\mu \phi  } }}{ e }^{ -\frac { 1 }{ \phi  } \left[ \mathbb{W}\left( -\frac { \mu \phi  }{ 1+\mu \phi  } {{ S }_{ 0 }\left( t \right)}{ e }^{ -\frac { \mu \phi  }{ 1+\mu \phi  }  } \right) +\displaystyle\frac { \mu \phi  }{ 1+\mu \phi  }  \right]  }$ \\
\hline
$h(t)$ &  $-\displaystyle\frac {  \omega h_0 \left( t \right)  { e }^{ -\frac { 1 }{ \phi  } \left[ \mathbb{W}\left( -\frac { \mu \phi  }{ 1+\mu \phi  } { S }_{ 0 }\left( t \right) { e }^{ -\frac { \mu \phi  }{ 1+\mu \phi  }  } \right) +\frac { \mu \phi  }{ 1+\mu \phi  }  \right]  } }{ \left( 1-e^{-\frac { \mu  }{ 1+\mu \phi  } } \right) \phi S\left(t \right)  } \frac {\mathbb{W}\left( -\frac { \mu \phi  }{ 1+\mu \phi  } { S }_{ 0 }\left( t \right) { e }^{ -\frac { \mu \phi  }{ 1+\mu \phi  }  } \right)  }{ 1+\mathbb{W}\left( -\frac { \mu \phi  }{ 1+\mu \phi  } { S }_{ 0 }\left( t \right) { e }^{ -\frac { \mu \phi  }{ 1+\mu \phi  }  } \right)  } $ \\
\hline
$f(t)$ & $-\displaystyle\frac {  \omega h_0 \left( t \right)  { e }^{ -\frac { 1 }{ \phi  } \left[ \mathbb{W}\left( -\frac { \mu \phi  }{ 1+\mu \phi  } { S }_{ 0 }\left( t \right) { e }^{ -\frac { \mu \phi  }{ 1+\mu \phi  }  } \right) +\frac { \mu \phi  }{ 1+\mu \phi  }  \right]  } }{ \left( 1-e^{-\frac { \mu }{ 1+\mu \phi  } } \right) \phi  } \frac {\mathbb{W}\left( -\frac { \mu \phi  }{ 1+\mu \phi  } { S }_{ 0 }\left( t \right) { e }^{ -\frac { \mu \phi  }{ 1+\mu \phi  }  } \right)  }{ 1+\mathbb{W}\left( -\frac { \mu \phi  }{ 1+\mu \phi  } { S }_{ 0 }\left( t \right) { e }^{ -\frac { \mu \phi  }{ 1+\mu \phi  }  } \right)  } $  \\
\hline
$q_0$ & $1-\omega$ \\
\hline
\end{tabular}
\end{table}

\section{Bayesian inference for the HZMGP-based frailty survival model} \label{Section:bayesian_inference}

Bayesian inference involves the combination of two sources of information: the data, represented through the likelihood function, and prior beliefs about the parameters, encoded in prior distributions. These are combined via Bayes' theorem to yield the posterior distribution, which reflects the updated uncertainty about the parameters after observing the data.\cite{gelman2013bayesian} We describe each of these components in the subsections below: the specification of the likelihood function, the choice of prior distributions, and the computational strategy for obtaining samples from the posterior distribution.

\subsection{Likelihood function} \label{Subsection:like}

Let $T > 0$ denote the random variable time until the occurrence of the event of interest, and let $\delta \in \{0,1\}$ be the censoring indicator, where $\delta = 1$ if the event is observed and $\delta = 0$ otherwise. We consider a sample of $n$ independent individuals, indexed by $i = 1,\ldots,n$, with data denoted as $\mathcal{D} = \left({\bm t}, {\bm\delta}, {\bm X} \right)$, where ${\bm t} = \left(t_1, \ldots, t_n \right)^\top$ are the lifetime of the subjects in the study, ${\bm \delta} = \left(\delta_1, \ldots, \delta_n \right)^\top$ are the indicators of censorship, and ${\bm X}$ is the matrix of covariates. 
We assume that survival times $T_1, \ldots, T_n$ are independent of censoring times. Hence, we assume that $T_i \sim \text{HZMGP}(\omega_i, \mu_i, \phi, \lambda, \gamma)$, where individual-specific parameters $\omega_i$ and $\mu_i$ are linked to covariates via logarithmic and logit link functions, respectively:
\begin{equation} \label{regressao_mu}
\omega_i = \frac{e^{\bm{x}_{i}^\top {\bm\beta}_p^{(\omega)}}}{1+e^{\bm{x}_{i}^\top{\bm\beta}_p^{(\omega)}}} \quad \text{and} \quad
\mu_i = e^{\bm{x}_{i}^\top {\bm\beta}_p^{(\mu)}},
\end{equation}

\noindent where $\bm{x}_{i}^\top= \left(1, { x }_{ i1 },\ldots,{ x }_{ ip } \right) $ is the $p \times 1$ vector of covariates for the $i$-th individual, ${\bm\beta}^{(\omega)}$ and ${\bm\beta}^{(\mu)}$
denote the vectors of regression coefficients. Then, the likelihood function for ${\bm\vartheta} = ({\bm\beta}^{(\omega)}, {\bm\beta}^{(\mu)}, \phi, \lambda, \gamma)^\top$ is described by
\begin{align}
\mathcal{L} \left( \mathcal{D} \mid {\bm\vartheta}  \right) %&= \prod _{ i=1 }^{ n }{ S\left( { t }_{ i }\right) \left[ { h\left( { t }_{ i } \right)} \right]^{ { \delta  }_{ i } } } \\
&= \prod _{ i=1 }^{ n } {\left[  1-\displaystyle\frac{\omega_i}{1-e^{-\frac { \mu_i}{ 1+\mu_i \phi  } }}+\displaystyle\frac{\omega_i}{1-e^{-\frac { \mu_i  }{ 1+\mu_i \phi  } }}{ e }^{ -\frac { 1 }{ \phi  } \left[ \mathbb{W}\left( -\frac { \mu_i \phi  }{ 1+\mu_i \phi  } {{ S }_{ 0 }\left( t_i \right)}{ e }^{ -\frac { \mu_i \phi  }{ 1+\mu_i \phi  }  } \right) +\displaystyle\frac { \mu_i \phi  }{ 1+\mu_i \phi  }  \right]  } \right]} \nonumber \\
& \times { {\left[ -\displaystyle\frac {  \omega_i h_0 \left( t_i \right)  { e }^{ -\frac { 1 }{ \phi  } \left[ \mathbb{W}\left( -\frac { \mu \phi  }{ 1+\mu \phi  } { S }_{ 0 }\left( t_i \right) { e }^{ -\frac { \mu_i \phi  }{ 1+\mu_i \phi  }  } \right) +\frac { \mu_i \phi  }{ 1+\mu_i \phi  }  \right]  } }{ \left( 1-e^{-\frac { \mu_i  }{ 1+\mu_i \phi  } } \right) \phi S\left(t_i \right)  } \frac {\mathbb{W}\left( -\frac { \mu_i \phi  }{ 1+\mu_i \phi  } { S }_{ 0 }\left( t_i \right) { e }^{ -\frac { \mu_i \phi  }{ 1+\mu_i \phi  }  } \right)  }{ 1+\mathbb{W}\left( -\frac { \mu_i \phi  }{ 1+\mu_i \phi  } { S }_{ 0 }\left( t_i \right) { e }^{ -\frac { \mu_i \phi  }{ 1+\mu_i \phi  }  } \right)  } \right]}^{\delta_i} }.
\label{vero_zmgp}
\end{align}

\subsection{Prior elicitation} \label{Subsection:priors}

We assign prior distributions to the model parameters to complete the Bayesian specification. Let  ${\bm\vartheta} = ( {\bm\beta}^{(\omega)}, {\bm\beta}^{(\mu)}, \theta_\phi, \theta_\lambda, \theta_\gamma)^\top$, where the transformed parameters are defined as
\begin{equation}
\phi = e^{\theta_\phi}, \quad \lambda = e^{\theta_\lambda}, \quad \gamma = e^{\theta_\gamma}.
\end{equation}

This log-transformation ensures that these parameters remain strictly positive while enabling a more symmetric and numerically stable parameter space for Bayesian computation.\cite{alvares2024bayesian} Assuming prior independence among all components, the joint prior distribution can be factorized as
\begin{equation}
\pi({\bm\vartheta}) = \pi( {\bm\beta}^{(\omega)}, {\bm\beta}^{(\mu)}, \theta_{\phi}, \theta_{\lambda}, \theta_{\gamma}) = \pi({\bm\beta}^{(\omega)}) \, \pi({\bm\beta}^{(\mu)}) \, \pi(\theta_{\phi}) \, \pi(\theta_{\lambda}) \, \pi(\theta_{\gamma}).
\label{priori}
\end{equation}

We assume weakly informative priors for all parameters. Specifically, we set a normal distribution with mean zero and variance $10^2$ for each component of the vector ${\bm\vartheta}$. So, the full Bayesian model can then be expressed in hierarchical form as follows:
\begin{equation} \label{Eq:Bayesian_model}
\begin{aligned} 
(t_i, \delta_i \mid \omega_i, \mu_i, \theta_\phi, \theta_\lambda, \theta_\gamma) &\sim \text{HZMGP}(\omega_i, \mu_i, \theta_\phi, \theta_\lambda, \theta_\gamma), \quad i = 1, \dots, n,\\
\omega_i &= \frac{e^{\bm{x}_{i}^\top {\bm\beta}^{(\omega)}}}{1+e^{\bm{x}_{i}^\top {\bm\beta}^{(\omega)} }}, \\
\mu_i &= e^{\bm{x}_{i}^\top {\bm\beta}^{(\mu)}},\\
{\bm\beta}^{(\omega)}, \; {\bm\beta}^{(\mu)}, \; \theta_\phi, \; \theta_\lambda, \; \theta_\gamma &\sim \mathcal{N}(0, \, 10^2).
\end{aligned}
\end{equation}

This hierarchical formulation summarizes the full model structure and serves as the basis for posterior inference via Hamiltonian Monte Carlo methods.\cite{neal2011mcmc}

\subsection{Posterior distribution}

In the Bayesian framework, inference is based on the posterior distribution, which synthesizes prior information and observed data to provide a complete probabilistic characterization of the model parameters. Given the likelihood function defined in Section~\ref{Subsection:like} and the prior distributions specified in Section~\ref{Subsection:priors}, the joint posterior distribution is expressed as
\begin{equation}
\pi({\bm\vartheta} \mid \mathcal{D}) \propto \mathcal{L}(\mathcal{D} \mid {\bm\vartheta}) \, \pi({\bm\beta}^{(\omega)}) \, \pi({\bm\beta}^{(\mu)}) \, \pi(\theta_{\phi}) \, \pi(\theta_{\lambda}) \, \pi(\theta_{\gamma}).
\label{posteriori}
\end{equation}

Due to the complexity of the likelihood function, particularly the nonlinearity introduced by the Lambert W function and the log-link structures for $\omega$ and $\mu$, the posterior distribution denoted in Equation \eqref{posteriori} is highly non-standard and lacks a closed-form solution. This renders direct analytical inference infeasible and requires efficient numerical methods. Hence, to perform Bayesian inference, we used Stan,\cite{carpenter2017stan} which allows flexible specification of complex hierarchical models and supports efficient gradient-based inference methods. 

Stan implements the No-U-Turn sampler, a state-of-the-art adaptive version of Hamiltonian Monte Carlo (HMC), which is particularly effective in exploring high-dimensional and correlated posterior distributions.\cite{neal2011mcmc, hoffman2014no} HMC generates proposals by simulating Hamiltonian dynamics using the gradient of the log-posterior density, allowing efficient transitions that reduce autocorrelation and avoid the random walk behavior of traditional MCMC. Stan internally performs automatic differentiation to compute exact gradients, which is essential in our context due to the non-standard likelihood induced by the HZMGP distribution. In addition, Stan's diagnostics, including the potential scale reduction factor ($\hat{R}$) and effective sample size (ESS), provide robust tools to assess convergence and efficiency.\cite{carpenter2017stan}

\subsection{Posterior-based classification of the zero-modification specification} \label{Subsection:posterior_classification}

Once posterior distribution samples are obtained, probabilistic classification of individuals into distinct zero-modification specifications becomes possible. In particular, the posterior distributions of the parameters $\omega_i$, $\mu_i$ and $\phi$ are essential to identify the structural form of the HZMGP distribution that best describes each individual. Recalling the theoretical characterization of the HZMPS distribution shown in Figure \ref{Figure:HZMPS}, the classification is governed by the relationship between the individual-specific value of $\omega_i$ and the threshold function 
\begin{equation}
\mbox{Threshold}(\mu_i,\phi) = 1 - \pi_{_\text{PS}}(Y_i=0; \, \mu_i,\phi)= 1 - e^{\left(-\displaystyle\frac{\mu_i}{1 + \mu_i \phi} \right)}. \label{threshold}
\end{equation}

This threshold separates the different subfamilies of the HZMGP distribution. Given that $\omega_i$, $\mu_i$, and $\phi$ are inferred via their joint posterior distributions, we can assess the classification of each individual by calculating the posterior probabilities of the inequality conditions. Specifically, we use the following Bayesian decision rule, based on the posterior probability thresholds ($1-\alpha$) to assign each patient to a zero-modification specification:
\begin{equation}
\begin{cases} 
\mathbb{P} \left[ \omega_i < 1-e^{\left(-\displaystyle\frac{\mu_i}{1 + \mu_i \phi} \right)}  \right] \ge 1- \alpha \quad \text{and} \quad \mathbb{P}[\omega_i < 1] \ge 1- \alpha & \text{ZIGP}, \\
\alpha < \mathbb{P} \left[ \omega_i < 1-e^{\left(-\displaystyle\frac{\mu_i}{1 + \mu_i \phi} \right)}  \right] < 1- \alpha \quad \text{and} \quad \mathbb{P}[\omega_i < 1] \ge 1- \alpha & \text{GP}, \\
\mathbb{P} \left[ \omega_i > 1-e^{\left(-\displaystyle\frac{\mu_i}{1 + \mu_i \phi} \right)} \right] \ge 1- \alpha \quad \text{and} \quad \mathbb{P}[\omega_i < 1] \ge 1- \alpha & \text{ZDGP}, \\
\mathbb{P}[ \omega_i \geq 1] \ge 1- \alpha & \text{ZTGP}.
\end{cases}
\label{bay_casos}
\end{equation}

This classification approach allows the model to identify the zero-modification structure that best fits each patient's posterior configuration, providing interpretable conclusions and enabling stratification of individuals based on their survival profiles. Importantly, the results of this probabilistic classification can be summarized both at the individual and population levels, offering a powerful descriptive layer to the proposed model.
%%%%%%%%%%%%%%%%%%%%%%%%%%%%%%%%%%%%%%%

%%%%%%%%%%%%%%%%%%%%%%%%%%%%%%%%%%%%%%%
\section{Simulation study} \label{Section:simulation}

To assess the performance of the proposed model, we performed a simulation study. The main goal was to examine the model's ability to accurately recover true parameter values across different scenarios and sample sizes. Additionally, we also evaluated how the model classifies each patient in different zero-modification specifications. 

We explored two simulated scenarios from the HZMGP-based model (see Section \ref{Section:bayesian_inference}), in which each of them exemplifies a different situation of proportion of zeros. Both scenarios consider two randomly generated covariates from a normal and a Bernoulli distribution. Each scenario is composed of 100 datasets with three sample sizes $n=1000, 10000, 20000$, and its parameters are specified as follows:

{\bf Scenario I:} $\phi = 0.25$, $\beta_{0}^{(\omega)} = 0.90$, $\beta_{1}^{(\omega)} = 0.23$, $\beta_{2}^{(\omega)} = -1.40$, $\beta_{0}^{(\mu)}=3.90$, $\beta_{1}^{(\mu)}= 0.13$, $\beta_{2}^{(\mu)} = -2.40$, $\lambda = 0.04$, $\gamma = 1.30$.

{\bf Scenario II:} $\phi = 0.13$, $\beta_{0}^{(\omega)} = 1.50$, $\beta_{1}^{(\omega)} = 0.60$, $\beta_{2}^{(\omega)} = 3.50$, $\beta_{0}^{(\mu)}=1.40$, $\beta_{1}^{(\mu)}= 0.10$, $\beta_{2}^{(\mu)} = 1.20$, $\lambda = 0.11$, $\gamma = 1.08$.

In each simulation, we fit the hierarchical model \eqref{Eq:Bayesian_model} considering three Markov chains, each having 1000 iterations and a burn-in period of 300. Table \ref{Table:simu_bayes} summarizes the results in terms of posterior mean, standard deviation (SD), 95\% coverage probability (CP), and the proportion of zero-modification specifications with threshold $\alpha = 0.1$ (see Section \ref{Subsection:posterior_classification}).

\begin{table}[ht]
\centering
\caption{Results of the simulation studies for the two scenarios.}
\label{Table:simu_bayes}
\begin{tabular}{cl|cc|cc|cc}
\cline{2-8}
& \multirow{2}{*}{Parameter} & \multicolumn{2}{|c|}{$n = 1000$} & \multicolumn{2}{|c|}{$n = 10000$} &  \multicolumn{2}{|c}{$n = 20000$}  \\
\cline{3-8}
&  & Mean (SD) & CP & Mean (SD) & CP & Mean (SD) & CP \\ 
\hline
\multicolumn{1}{c|}{\multirow{13}{*}{\begin{sideways}Scenario I\end{sideways}}} & $\phi = 0.25$ & 0.371 (0.560) & 0.95 & 0.214 (0.028) & 0.99 & 0.217 (0.027) & 0.99\\
\multicolumn{1}{c|}{} & $\beta_{0}^{(\omega)}=0.90$ & 0.931 (0.187) & 0.96  & 0.901 (0.064) & 0.95 &0.905 (0.050) & 0.93 \\ 
\multicolumn{1}{c|}{} & $\beta_{1}^{(\omega)}=0.23$ & 0.234 (0.064) & 0.95  & 0.229 (0.022) & 0.98 &0.229 (0.015) & 0.96 \\ 
\multicolumn{1}{c|}{} & $\beta_{2}^{(\omega)}=-1.40$ & -1.440 (0.209) & 0.95  & -1.397 (0.070) & 0.96 &-1.401 (0.053) & 0.93 \\ 
\multicolumn{1}{c|}{} & $\beta_{0}^{(\mu)}=3.90$ & 4.168 (2.244) & 0.74  & 4.285 (0.349) & 0.99 &4.220 (0.329) & 0.98 \\ 
\multicolumn{1}{c|}{} & $\beta_{1}^{(\mu)}=0.13$ &  0.144 (0.198) & 0.91 & 0.127 (0.037) & 0.95 &0.125 (0.024) & 0.95 \\ 
\multicolumn{1}{c|}{} & $\beta_{2}^{(\mu)}=-2.40$ & -5.385 (2.593) & 0.90  & -2.368 (0.173) & 0.96 &-2.361 (0.132) & 0.91 \\
\multicolumn{1}{c|}{} & $\lambda = 0.04$ &  0.095 (0.068) & 0.80  & 0.033 (0.010) & 0.99 &0.033 (0.008) & 0.99\\
\multicolumn{1}{c|}{} & $\gamma=1.30$ &  1.211 (0.143) & 0.72  & 1.285 (0.030) & 0.91 & 1.290 (0.021) & 0.94 \\
\cline{2-8}
& \multicolumn{1}{|c|}{\multirow{4}{*}{Classification}} & ZIGP & 0.49 & ZIGP & 0.99 & ZIGP & 1.00 \\
\multicolumn{1}{c|}{} &  & GP & 0.46 & GP & 0.01 & GP & 0.00\\
\multicolumn{1}{c|}{} &  & ZDGP & 0.05 & ZDGP & 0.00 & ZDGP & 0.00\\
\multicolumn{1}{c|}{} &  & ZTGP & 0.00 & ZTGP & 0.00 & ZTGP & 0.00 \\
\hline
\multicolumn{1}{c|}{\multirow{13}{*}{\begin{sideways}Scenario II\end{sideways}}} & $\phi = 0.13$ & 0.116 (0.039) & 0.99 & 0.130 (0.011) & 0.98 & 0.131 (0.008) & 0.97 \\
\multicolumn{1}{c|}{} & $\beta_{0}^{(\omega)}=1.50$ & 1.536 (0.260) & 0.93 & 1.516 (0.075) & 0.93 & 1.511 (0.046) & 0.96 \\ 
\multicolumn{1}{c|}{} & $\beta_{1}^{(\omega)}=0.60$ & 0.595 (0.234) & 0.92 & 0.597 (0.069) & 0.93 & 0.603 (0.050) & 0.94 \\ 
\multicolumn{1}{c|}{} & $\beta_{2}^{(\omega)}=3.50$ & 4.441 (1.552) & 0.93 & 3.525 (0.160) & 0.97 & 3.525 (0.126) & 0.93\\ 
\multicolumn{1}{c|}{} & $\beta_{0}^{(\mu)}=1.40$ & 1.719 (0.972) & 1.00 & 1.623 (0.266) & 0.99 & 1.612 (0.242) & 0.94\\ 
\multicolumn{1}{c|}{} & $\beta_{1}^{(\mu)}=0.10$ & 0.118 (0.063) & 0.96 & 0.107 (0.017) & 0.95 & 0.105 (0.012) & 0.95\\ 
\multicolumn{1}{c|}{} & $\beta_{2}^{(\mu)}=1.20$ & 1.665 (0.471) & 0.90 & 1.246 (0.060) & 0.93 & 1.235 (0.041) & 0.91\\
\multicolumn{1}{c|}{} & $\lambda = 0.11$ & 0.101 (0.042) & 1.00 & 0.087 (0.019) & 0.99  & 0.085 (0.018) & 0.93\\
\multicolumn{1}{c|}{} & $\gamma=1.08$ & 1.118 (0.071) & 0.73 & 1.099 (0.022) & 0.94 & 1.097 (0.016) & 0.95 \\
\cline{2-8}
& \multicolumn{1}{|c|}{\multirow{4}{*}{Classification}} & ZIGP & 0.03 & ZIGP & 0.15 & ZIGP & 0.29 \\
\multicolumn{1}{c|}{} &  & GP & 0.97 & GP & 0.84 & GP & 0.69\\
\multicolumn{1}{c|}{} &  & ZDGP & 0.00 & ZDGP & 0.01 & ZDGP & 0.02\\
\multicolumn{1}{c|}{} &  & ZTGP & 0.00 & ZTGP & 0.00 & ZTGP & 0.00 \\
\hline
\end{tabular}
\end{table}

As expected, the posterior mean approaches the true parameter value and the standard deviation decreases as the sample size increases. The 95\% coverage probability is robust for large sample sizes ($n=10000,20000$) and presents a medium coverage ($\sim$70\%) for a few parameters when $n=1000$. The proportion of zero-modification specifications also stabilizes as the sample increases. For Scenario I with $n=20000$, all frailty terms are induced by a ZIGP distribution, which suggests a high cure rate. For Scenario II with $n=20000$, the frailty terms with zero inflation reduce to 29\%, while only 2\% indicates zero deflation and the remainder is characterized by a (traditional) generalized Poisson distribution. It is worth noting that the ZDGP classification may indicate that the individual is at accelerated risk, which, in an applied context, suggests some (medical) intervention.
%%%%%%%%%%%%%%%%%%%%%%%%%%%%%%%%%%%%%%%

%%%%%%%%%%%%%%%%%%%%%%%%%%%%%%%%%%%%%%%
\section{Case study: lung cancer survival analysis} \label{Section:application}

According to the Brazilian National Cancer Institute\cite{inca2023}, approximately 18000 men were diagnosed with lung cancer in 2023, making it the third most common type of cancer in this group. In women, lung cancer is the fourth most prevalent, with approximately 8000 new cases recorded in 2023. This reflects the high incidence of this disease, which makes it one of the main causes of morbidity and mortality in Brazil.

In this study, we used lung cancer data provided by the S{\~a}o Paulo Oncocenter Foundation, a public institution affiliated with the S{\~a}o Paulo State Health Secretariat, Brazil. The dataset is available at \url{https://fosp.saude.sp.gov.br/}. These data were initially analyzed by Gazon et al.\cite{gazon2022nonproportional}, who examined the effects of several covariates: age (categorized as younger than 60 or older than 60), clinical stage (divided into four groups), and treatment variables such as surgery, radiotherapy, and chemotherapy (each classified as received or not). However, in this paper, age was analyzed as a continuous variable and the clinical stage was reclassified into two categories: I-II and III-IV.

The data is composed of 30900 patients residing in the state of S{\~a}o Paulo and diagnosed with lung cancer between 2000 and 2014, and followed up until 2018. Table \ref{Table:dados} describes the distribution of each covariate.

\begin{table}[ht]
\centering
\caption{Descriptive analysis of covariates in the lung cancer data.}
\label{Table:dados}
\begin{tabular}{l|c|c|c|c}
\hline
\multicolumn{1}{c|}{Covariate} & Category & Description &  $n$ & $\%$ \\ 
\hline
\multirow{2}{*}{Age: Age (in years)}  & \multirow{2}{*}{--} & $\mbox{Mean}=63.21$  & \multirow{2}{*}{$30900$} & \multirow{2}{*}{--} \\
&  & $\mbox{SD}=10.98$ & \\
\hline
\multirow{2}{*}{Gen: Gender} & $0$ & {Male} & $19657$ & $63.61 \% $   \\
 & $1$ & {Female} & $11243$ & $36.39 \%$ \\
\hline
\multirow{2}{*}{CS: Clinical stage}
 & $0$ & I or II & $4929$ & $15.95\% $ \\
& $1$ & III or IV & $25971$ & $84.05 \% $ \\
\hline
\multirow{2}{*}{Sur: Surgery} & $0 $ & No & $25001$ & $80.91\% $ \\
& $1$ & Yes & $5899$ & $19.09\% $ \\
\hline
\multirow{2}{*}{Rad: Radiotherapy} & $0 $ & No & $19109 $ & $61.84\% $ \\
& $1$ & Yes & $11791 $ & $38.16\% $ \\
\hline
\multirow{2}{*}{Che: Chemotherapy}  & $0 $ & No & $10766$ & $34.84\% $ \\
& $1 $ & Yes & $20134$ & $65.16\% $ \\
\hline
\end{tabular}
\end{table}

Our outcome of interest is the time to death from lung cancer. Due to the high level of aggressiveness of this disease, approximately 90\% of the patients die during the study. Although some patients survive for more than 15 years, the median survival time is less than one year. To deepen our understanding of these data, we propose the HZMGP-based model \eqref{Eq:Bayesian_model}, where the parameters $\omega_i$ (zero-modification probability) and $\mu_i$ (mean) are specified as follows:
\begin{equation} \label{Eq:Bayesian_model_lung}
\begin{aligned} 
\omega_i & =  \frac{e^{\eta^{(\omega)}_i}}{1 + e^{\eta^{(\omega)}_i}}, \\
\eta^{(\omega)}_i & =  \beta_{0}^{(\omega)} + \beta_{1}^{(\omega)} \text{Age}_i + \beta_{2}^{(\omega)} \text{Gen}_i + \beta_{3}^{(\omega)} \text{CS}_i + \beta_{4}^{(\omega)} \text{Sur}_i + \beta_{5}^{(\omega)} \text{Rad}_i + \beta_{6}^{(\omega)} \text{Che}_i, \\
\mu_i & =  e^{\eta^{(\mu)}_i},  \\
\eta^{(\mu)}_i & =  \beta_{0}^{(\mu)} + \beta_{1}^{(\mu)} \text{Age}_i + \beta_{2}^{(\mu)} \text{Gen}_i + \beta_{3}^{(\mu)} \text{CS}_i + \beta_{4}^{(\mu)} \text{Sur}_i + \beta_{5}^{(\mu)} \text{Rad}_i + \beta_{6}^{(\mu)} \text{Che}_i.
\end{aligned}
\end{equation}

As in the simulation study, we set three Markov chains, each having 1000 iterations and a burn-in period of 300. To assess convergence and efficiency, we examined both the potential scale reduction factor ($\hat{R}$) and the effective sample size (ESS). All parameters presented $\hat{R}$ values equal to 1, indicating that Monte Carlo chains had successfully converged. In addition, the ESS values were large enough to ensure a reliable inference ($>900$). Table \ref{Table:todascov} summarizes the results in terms of posterior mean, standard deviation (SD), 95\% credible interval, and the proportion of zero-modification specifications with threshold $\alpha = 0.1$ (see Section \ref{Subsection:posterior_classification}).

\begin{table}[ht]
\centering
\caption{Posterior summary of the model \eqref{Eq:Bayesian_model} fitted to the lung cancer data and the proportion of zero-modification specifications with threshold $\alpha = 0.1$.}
\label{Table:todascov}
\begin{tabular}{l|cccc}
\hline
\multicolumn{1}{c|}{\multirow{2}{*}{Parameter}}& & & \multicolumn{2}{c}{{Credible interval}}\\
\cline{4-5}
 & {Mean} & {SD} & {2.5\%} & {97.5\%} \\
\hline
$\phi$ \, [Dispersion]    & 0.140  & 0.003  & 0.134  & 0.147  \\
$\beta_{0}^{(\omega)}$ \, [Intercept] & 2.370  & 0.107  & 2.160  & 2.582  \\
$\beta_{1}^{(\omega)}$ \, [Age] & 0.454  & 0.039  & 0.379  & 0.528  \\
$\beta_{2}^{(\omega)}$ \, [Gen: F] & -0.603 & 0.078  & -0.755 & -0.450 \\
$\beta_{3}^{(\omega)}$ \, [CS: III-IV] & 2.322  & 0.098  & 2.136  & 2.511  \\
$\beta_{4}^{(\omega)}$ \, [Sur: Y] & -1.964 & 0.097  & -2.149 & -1.773 \\
$\beta_{5}^{(\omega)}$ \, [Rad: Y] & 0.361  & 0.108  & 0.153  & 0.570  \\
$\beta_{6}^{(\omega)}$ \, [Che: Y] & 0.744  & 0.091  & 0.567  & 0.918  \\
$\beta_{0}^{(\mu)}$ \, [Intercept] & 2.402  & 0.070  & 2.267  & 2.536  \\
$\beta_{1}^{(\mu)}$ \, [Age] & -0.061 & 0.011  & -0.082 & -0.039 \\
$\beta_{2}^{(\mu)}$ \, [Gen: F] & -0.267 & 0.022  & -0.312 & -0.224 \\
$\beta_{3}^{(\mu)}$ \, [CS: III-IV] & 1.666  & 0.054  & 1.563  & 1.780  \\
$\beta_{4}^{(\mu)}$ \, [Sur: Y] & -1.291 & 0.043  & -1.375 & -1.207 \\
$\beta_{5}^{(\mu)}$ \, [Rad: Y] & -0.658 & 0.024  & -0.704 & -0.611 \\
$\beta_{6}^{(\mu)}$ \, [Che: Y] & -1.936 & 0.032  & -2.002 & -1.874 \\
$\lambda$ \, [Weibull scale]  & 0.187  & 0.007  & 0.174  & 0.201  \\
$\gamma$ \, [Weibull shape]    & 1.266  & 0.007  & 1.252  & 1.280  \\
\hline
\multicolumn{1}{c|}{\multirow{2}{*}{Classification}} & ZIGP  & GP & ZDGP & ZTGP \\
\cline{2-5}
& 0.32 & 0.03 & 0.65 & 0.00 \\
\hline
\end{tabular}
\end{table}

The posterior distribution of $\phi$ suggests the presence of unmeasured risk factors that influence patient mortality. We hypothesize that this unmeasured variability primarily stems from underlying biological differences, such as tumor molecular subtypes or unassessed aspects of the patient's general condition. This interpretation, which biologically contextualizes the frailty component by linking it to these specific factors, is robustly supported by recent seminal multiomic studies.\cite{cancer2014comprehensive, zhang2024multi, wang2025multi} These investigations conclusively demonstrate the existence of molecular subtypes in lung cancer (e.g., lung adenocarcinoma) with intrinsically distinct prognoses and differential responses to standard therapies such as chemotherapy and immunotherapy.\cite{zhang2024multi}

Also in Table \ref{Table:todascov}, all covariates contribute significantly to explaining the zero-modification probability $\omega_i$. The same happens for the mean parameter $\mu_i$, where the risk of death increases in young men with clinical stage III or IV, and without treatment (no surgery, no radiotherapy, and no chemotherapy). This conclusion coincides with the findings of Gazon et al.\cite{gazon2022nonproportional}, who studied this same sample of patients.

Finally, the proportion of zero-modification specifications is calculated according to \eqref{bay_casos}, where the structural form of the HZMGP distribution that best describes each patient can be classified as zero-inflated (ZIGP), zero-deflated (ZDGP), zero-truncated (ZTGP), or traditional generalized Poisson (GP) distribution. Here, the majority of patients (65\%) were classified in the ZDGP group. This means that they probably had a poor prognosis and the need for intensive clinical follow-up to recover from lung cancer. In contrast, 32\% of the patients presented characteristics consistent with a higher probability of recovery (i.e., zero inflation), indicating a favorable profile for improvement. The remaining 3\% were assigned to the GP group, corresponding to a standard-risk profile, for which adherence to conventional lung cancer treatment guidelines is recommended. The computational time to fit the model was approximately three hours.

\begin{figure}[ht!]
\centering
\includegraphics[scale=.58]{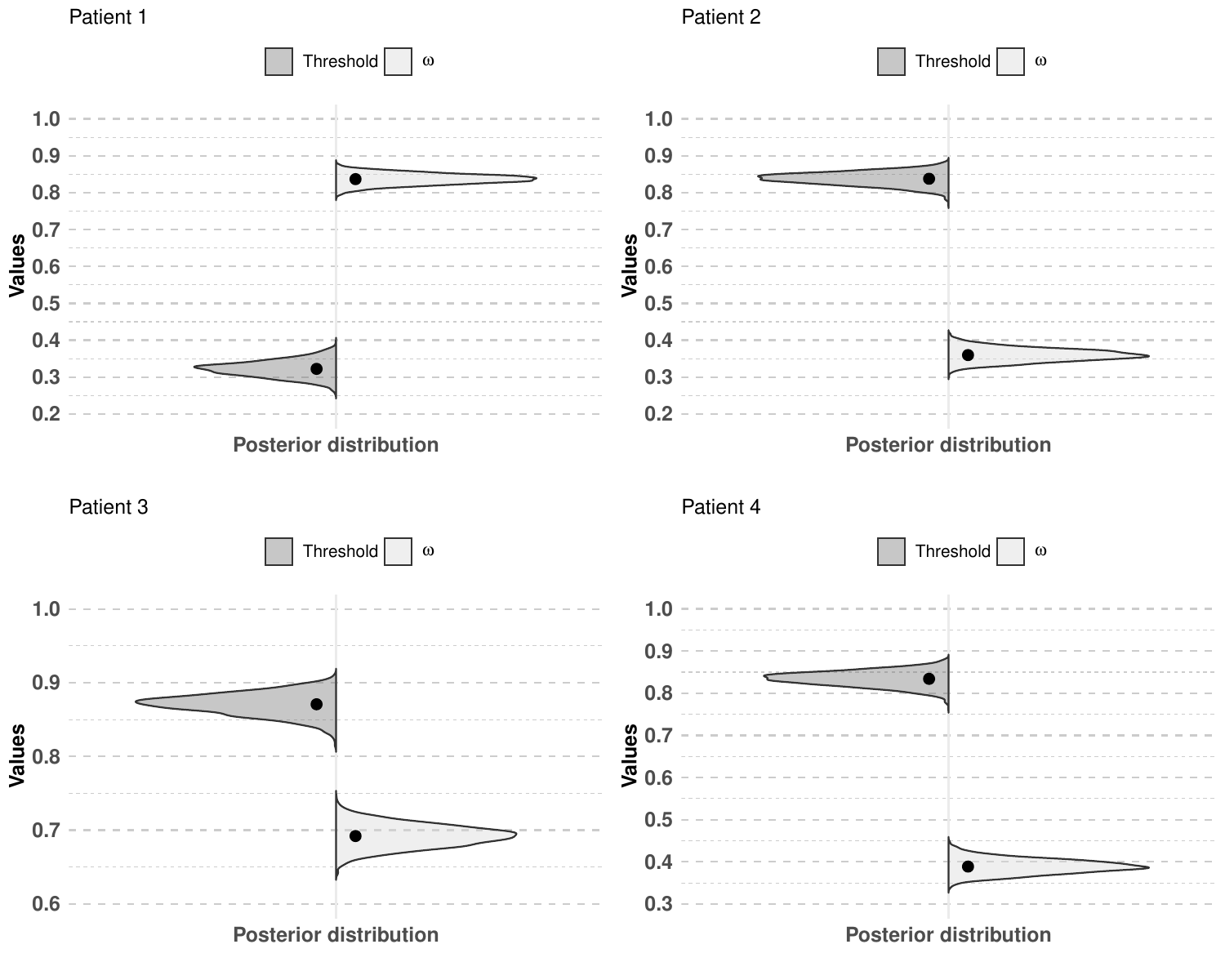}
\caption{Classification of zero-modification specification for the first four patients in the lung cancer data.}
\label{violin_plot_patients}
\end{figure}

Figure \ref{violin_plot_patients} illustrates the decision rule (see Figure \ref{Figure:HZMPS} and Section \ref{Subsection:posterior_classification} for more details) to classify the zero-modification specification for the first four patients in the lung cancer study.

Patient 1 has his/her parameter $\omega_i$ clearly greater than the threshold \eqref{threshold}, indicating that a ZDGP specification better captures the behavior of this patient. On the other hand, patients 2, 3 and 4 have their parameters $\omega_i$ lower than their respective thresholds, which favors the ZIGP specification.
%%%%%%%%%%%%%%%%%%%%%%%%%%%%%%%%%%%%%%%

%%%%%%%%%%%%%%%%%%%%%%%%%%%%%%%%%%%%%%%
\section{Final remarks} \label{Section:finalremarks}

This work makes a significant contribution to the existing literature by proposing a Bayesian survival model that incorporates the HZMGP distribution to model the frailty term. The main advantage of using this distribution lies in its inherent flexibility, which allows for accommodating diverse zero-modification specifications without requiring a priori knowledge of the underlying behavior of the data. Furthermore, the model's flexibility is enhanced by the inclusion of a dispersion parameter, allowing for more robust modeling under different scenarios of data variability.

In our simulated and applied studies, we illustrated the model's ability to perform individualized patient classification stands out, distinguishing them based on the zero-modification specification (zero-inflated, zero-deflated, zero-truncated, or traditional power series distribution). This classification is crucial to better identify patients with different risk profiles, which in turn enables the design of personalized follow-up strategies. In addition, the biological contextualization provided for $\phi$ (dispersion) transcends the statistical description, enriching the understanding of inter-individual variability. For example, the interpretation of this parameter allows for the formulation of hypotheses about the influence of underlying factors, such as specific genetic profiles or other comorbidities not explicitly modeled, that could modulate an individual's propensity for certain outcomes.

It should be noted that our simulation study showed that relatively small sample sizes ($n \leq 1000$) may bias the estimation of some parameters. Moreover, the computational time required for model fitting with lung cancer data is substantial (around 3 hours), a direct consequence of the inherent complexity of the proposed Bayesian structure and simulation-based estimation methods. To address these challenges, particularly the computational cost, we suggest the exploration of alternative computational tools as a future research direction. Specifically, approximation methods such as the integrated nested Laplace approximation (INLA)\cite{rue2009} could offer a promising avenue to perform inference in this type of model more efficiently, thus mitigating some of the current limitations.
%%%%%%%%%%%%%%%%%%%%%%%%%%%%%%%%%%%%%%%

\bibliographystyle{SageV}
\bibliography{references}

\begin{thebibliography}{10}
\providecommand{\url}[1]{\texttt{#1}}
\providecommand{\urlprefix}{URL }
\expandafter\ifx\csname urlstyle\endcsname\relax
  \providecommand{\doi}[1]{DOI:\discretionary{}{}{}#1}\else
  \providecommand{\doi}{DOI:\discretionary{}{}{}\begingroup \urlstyle{rm}\Url}\fi
\providecommand{\eprint}[2][]{\url{#2}}

\bibitem{Collett2015}
Collett D.
\newblock \emph{Modelling survival data in medical research}.
\newblock 3rd ed. Boca Raton, FL: Chapman and Hall/CRC, 2015.

\bibitem{Meeker2022}
Meeker WQ, Escobar LA and Pascual FG.
\newblock \emph{Statistical methods for reliability data}.
\newblock 2nd ed. Hoboken, NJ: John Wiley \& Sons, 2022.

\bibitem{Dickson2020}
Dickson DCM, Hardy MR and Waters HR.
\newblock \emph{Actuarial mathematics for life contingent risks}.
\newblock 3rd ed. Cambridge: Cambridge University Press, 2020.

\bibitem{Wang2019}
Wang P, Li Y and Reddy CK.
\newblock Machine learning for survival analysis: {A} survey.
\newblock \emph{ACM Computing Surveys (CSUR)} 2019; 51(6): 1--36.

\bibitem{WHO2024Cancer}
{World~Health~Organization}.
\newblock Cancer.
\newblock Fact sheet, 2024.
\newblock \urlprefix\url{https://www.who.int/news-room/fact-sheets/detail/cancer}.

\bibitem{Siegel2024}
Siegel RL, Miller KD, Wagle NS et~al.
\newblock Cancer statistics, 2024.
\newblock \emph{CA: A Cancer Journal for Clinicians} 2024; 74(1): 12--49.

\bibitem{ferlay2024global}
Ferlay J, Ervik M, Lam F et~al.
\newblock Global cancer observatory: {C}ancer today.
\newblock \emph{Lyon, France: International Agency for Research on Cancer} 2024; 2018-2020.

\bibitem{vaupel1979impact}
Vaupel JW, Manton KG and Stallard E.
\newblock The impact of heterogeneity in individual frailty on the dynamics of mortality.
\newblock \emph{Demography} 1979; 16(3): 439--454.

\bibitem{cox_oakes}
Cox DR and Oakes D.
\newblock \emph{Analysis of survival data}.
\newblock 1st ed. Boca Raton, FL: Chapman and Hall/CRC, 1984.

\bibitem{wienke2010frailty}
Wienke A.
\newblock \emph{Frailty models in survival analysis}.
\newblock 1st ed. Boca Raton, FL: Chapman and Hall/CRC, 2010.

\bibitem{caroni2010}
Caroni C, Crowder M and Kimber A.
\newblock Proportional hazards models with discrete frailty.
\newblock \emph{Lifetime Data Analysis} 2010; 16(3): 374--384.

\bibitem{ata2013survival}
Ata N and {\"O}zel G.
\newblock Survival functions for the frailty models based on the discrete compound {P}oisson process.
\newblock \emph{Journal of Statistical Computation and Simulation} 2013; 83(11): 2105--2116.

\bibitem{de2017bayesian}
Souza D, Cancho VG, Rodrigues J et~al.
\newblock Bayesian cure rate models induced by frailty in survival analysis.
\newblock \emph{Statistical Methods in Medical Research} 2017; 26(5): 2011--2028.

\bibitem{cancho2019new}
Cancho VG, Macera MA, Suzuki AK et~al.
\newblock A new long-term survival model with dispersion induced by discrete frailty.
\newblock \emph{Lifetime Data Analysis} 2020; 26: 221--244.

\bibitem{molina2021survival}
Molina KC, Calsavara VF, Tomazella VD et~al.
\newblock Survival models induced by zero-modified power series discrete frailty: {A}pplication with a melanoma data set.
\newblock \emph{Statistical Methods in Medical Research} 2021; 30(8): 1874--1889.

\bibitem{cancho2021bayesian}
Cancho VG, Barriga GD, Cordeiro GM et~al.
\newblock Bayesian survival model induced by frailty for lifetime with long-term survivors.
\newblock \emph{Statistica Neerlandica} 2021; 75(3): 299--323.

\bibitem{do2022survival}
Espirito~Santo APJ, Cancho VG, Louzada F et~al.
\newblock A survival model for lifetime with long-term survivors and unobserved heterogeneity.
\newblock \emph{Brazilian Journal of Probability and Statistics} 2022; 36(4): 692--703.

\bibitem{tsodikov2003}
Tsodikov AD, Ibrahim JG and Yakovlev AY.
\newblock Estimating cure rates from survival data: {A}n alternative to two-component mixture models.
\newblock \emph{Journal of the American Statistical Association} 2003; 98(464): 1063--1078.

\bibitem{carpenter2017stan}
Carpenter B, Gelman A, Hoffman MD et~al.
\newblock Stan: {A} probabilistic programming language.
\newblock \emph{Journal of Statistical Software} 2017; 76: 1--32.

\bibitem{frome1983analysis}
Frome EL.
\newblock The analysis of rates using {P}oisson regression models.
\newblock \emph{Biometrics} 1983; 39(3): 665--674.

\bibitem{brown2002test}
Brown LD and Zhao LH.
\newblock A test for the {P}oisson distribution.
\newblock \emph{Sankhy{\=a}: The Indian Journal of Statistics, Series A} 2002; 64(3): 611--625.

\bibitem{hayat2014understanding}
Hayat MJ and Higgins M.
\newblock Understanding {P}oisson regression.
\newblock \emph{Journal of Nursing Education} 2014; 53(4): 207--215.

\bibitem{inouye2017review}
Inouye DI, Yang E, Allen GI et~al.
\newblock A review of multivariate distributions for count data derived from the {P}oisson distribution.
\newblock \emph{Wiley Interdisciplinary Reviews: Computational Statistics} 2017; 9(3): e1398.

\bibitem{kurnia2021analysis}
Kurnia A, Sadik K et~al.
\newblock Analysis of overdispersed count data by {P}oisson model.
\newblock \emph{European Journal of Molecular \& Clinical Medicine} 2021; 7(10): 1400--1409.

\bibitem{kamalja2018estimation}
Kamalja KK and Wagh YS.
\newblock Estimation in zero-inflated generalized {P}oisson distribution.
\newblock \emph{Journal of Data Science} 2018; 16(1): 183--206.

\bibitem{conceicao2017}
Concei{\c{c}}{\~a}o KS, Louzada F, Andrade M et~al.
\newblock Zero-modified power series distribution and its hurdle distribution version.
\newblock \emph{Journal of Statistical Computation and Simulation} 2017; 87(9): 1842--1862.

\bibitem{conceiccao2017biparametric}
Concei{\c{c}}{\~a}o KS, Tomazella V, Andrade MG et~al.
\newblock Biparametric zero-modified power series distributions: {B}ayesian analysis under a reference prior approach.
\newblock \emph{Communications in Statistics-Theory and Methods} 2017; 46(21): 10518--10536.

\bibitem{ambagaspitiya1994compound}
Ambagaspitiya RS and Balakrishnan N.
\newblock On the compound generalized {P}oisson distributions.
\newblock \emph{ASTIN Bulletin: the Journal of the IAA} 1994; 24(2): 255--263.

\bibitem{corless1996lambertw}
Corless RM, Gonnet GH, Hare DE et~al.
\newblock On the {L}ambert {W} function.
\newblock \emph{Advances in Computational Mathematics} 1996; 5(1): 329--359.

\bibitem{hougaard1984life}
Hougaard P.
\newblock Life table methods for heterogeneous populations: {D}istributions describing the heterogeneity.
\newblock \emph{Biometrika} 1984; 71(1): 75--83.

\bibitem{dossantos1995}
Santos DM, Davies RB and Francis B.
\newblock Nonparametric hazard versus nonparametric frailty distribution in modelling recurrence of breast cancer.
\newblock \emph{Journal of Statistical Planning and Inference} 1995; 47(1-2): 111--127.

\bibitem{calsavara_pvf}
Calsavara V, Rodrigues A, Tomazella V et~al.
\newblock Frailty models power variance function with cure fraction and latent risk factors negative binomial.
\newblock \emph{Communications in Statistics-Theory and Methods} 2017; 46: 9763--9776.

\bibitem{gelman2013bayesian}
Gelman A, Carlin JB, Stern HS et~al.
\newblock \emph{Bayesian data analysis}.
\newblock 3rd ed. Boca Raton, FL: Chapman and Hall/CRC, 2013.

\bibitem{alvares2024bayesian}
Alvares D, Van~Niekerk J, Krainski ET et~al.
\newblock Bayesian survival analysis with {INLA}.
\newblock \emph{Statistics in Medicine} 2024; 43(20): 3975--4010.

\bibitem{neal2011mcmc}
Neal RM.
\newblock {MCMC} using {H}amiltonian dynamics.
\newblock \emph{Handbook of Markov Chain Monte Carlo} 2011; 2(11): 1--51.

\bibitem{hoffman2014no}
Hoffman MD and Gelman A.
\newblock The {No-U-Turn} sampler: {A}daptively setting path lengths in {H}amiltonian {M}onte {C}arlo.
\newblock \emph{Journal of Machine Learning Research} 2014; 15(1): 1593--1623.

\bibitem{inca2023}
{Instituto Nacional de Câncer (INCA)}.
\newblock C{\^a}ncer de pulm{\~a}o, 2023.
\newblock \urlprefix\url{https://www.inca.gov.br/tipos-de-cancer/cancer-de-pulmao}.

\bibitem{gazon2022nonproportional}
Gazon AB, Milani EA, Mota AL et~al.
\newblock Nonproportional hazards model with a frailty term for modeling subgroups with evidence of long-term survivors: {A}pplication to a lung cancer dataset.
\newblock \emph{Biometrical Journal} 2022; 64(1): 105--130.

\bibitem{cancer2014comprehensive}
{Cancer Genome Atlas Research Network}.
\newblock Comprehensive molecular profiling of lung adenocarcinoma.
\newblock \emph{Nature} 2014; 511: 543--550.

\bibitem{zhang2024multi}
Zhang Y, Wang Y and Qian H.
\newblock Multi-omics characterization and machine learning of lung adenocarcinoma molecular subtypes to guide precise chemotherapy and immunotherapy.
\newblock \emph{Frontiers in Immunology} 2024; 15: 1497300.

\bibitem{wang2025multi}
Wang C, Li J, Chen J et~al.
\newblock Multi-omics analyses reveal biological and clinical insights in recurrent stage {I} non-small cell lung cancer.
\newblock \emph{Nature Communications} 2025; 16(1): 1--19.

\bibitem{rue2009}
Rue H, Martino S and Chopin N.
\newblock Approximate {B}ayesian inference for latent {G}aussian models by using integrated nested {L}aplace approximations.
\newblock \emph{Journal of the Royal Statistical Society: Series B (Statistical Methodology)} 2009; 71(2): 319--392.

\end{thebibliography}

\end{document}